\documentstyle[12pt]{article}

\parskip =\medskipamount
\def\be{\begin{equation}}
\def\ee{\end{equation}}
\def\disp{\displaystyle}
\def\R{\rm {\sf R\hspace{-3.05mm}I\hspace{2.1mm}}}
\def\Z{\rm {\sf Z\hspace{-3.15mm}Z\hspace{0.6mm}}}

\newcounter{fig}

\begin{document}

\begin{titlepage}

\begin{center}
{\large \bf CORRELATION FUNCTIONS FOR SOME \\ CONFORMAL THEORIES ON \\
RIEMANN SURFACES

}
\vspace{0.5in}

{\sc Michael Monastyrsky}

{\sl Institute of Theoretical and Experimental Physics, \\
117259 Moscow, Russia} \\ e-mail: {\it monastyrsky@vxitep.itep.ru}
\medskip

{\sc Sergei Nechaev}

{\sl Landau Institute for Theoretical Physics, \\
117940, Moscow, Russia} \\ e-mail: {\it nechaev@landau.ac.ru}

\end{center}

\vspace{1in}

\begin{abstract}
We discuss the geometrical connection between 2D conformal field theories,
random walks on hyperbolic Riemann surfaces and knot theory. For the wide
class of CFT's with monodromies being the discrete subgroups of
$SL(2,{\R})$ the determination of four--point correlation functions are
related to construction of topological invariants for random walks on
multipunctured Riemann surfaces.
\end{abstract}

\end{titlepage}

\section*{Introduction}

The behavior of systems consisting of chain-like objects can be modified
in an essential way by topological constraints. The topological problems
have been widely investigated in connection with quantum field, string
theories and quantum
Hall effect \cite{wil}, vortices in superfluid liquids ($^3$He, $^4$He)
\cite{mon} and thermodynamic properties of entangled polymers
\cite{nech,grosb} etc. In the last few years there has been much progress
in understanding the relationship between Chern--Simons conformal field
theory on the one hand and constructions of knot and link invariants on
the other hand (see, for general references \cite{int,witten}). However,
despite the general concepts have been well elaborated in the
field-theoretic context, their application to the related areas of
mathematics and physics has been highly limited.

In this note we indicate how the simple geometrical constructions
originating from the theory of representations of braid groups enable us to
unite such at first sight different objects like CFT, random walks on
hyperbolic surfaces and knot invariants. The main physical
application due to this approach is the evaluation of four--point
correlation functions in CFT on the Riemann surface associated with the
discrete subgroups of the group $SL(2,{\R})$.

\section{Random Walk on Punctured Plane and Conformal Field Theory}

We start with the random walk of length $L$ and the elementary
step $\ell$ ($\ell\equiv 1$) on the complex plane ${\cal C}_1=\{z|\;
z=x+iy\}$ with two punctures. Suppose the coordinates of these points
being $M_1$ ($z_1=(0,0)$) and $M_2$ ($z_2=(a,0)$) ($a\equiv 1$) (such
choice does not indicate the loss of generality). Consider two closed paths
on ${\cal C}_1$ and attribute the generators $g_1,\, g_2$ of some group $G$
to the windings around the points $M_1$ and $M_2$. if we move along the
path in the clockwise direction (we apply $g_1^{-1},\, g_2^{-1}$). The
question is: what is the probability $P(\mu,L)$ for the random walk of
length $L$ on the plane ${\cal C}_1$ to form a closed loop with the
primitive word written in terms of generators $\{g_1, g_2, g_1^{-1},
g_2^{-1}\}$ to have the length $\mu$.

The function $P(\mu,L)$ can be written as a path integral with a Wiener
measure
\be \label{eq:8}
P(\mu,L)=\frac{1}{\Theta} \int\ldots\int {\cal D}\{z\}
\exp\left\{-\int_0^L \left(\frac{dz(s)}{ds}\right)^2ds\right\}
\delta\left[W\{g_1, g_2, g_1^{-1}, g_2^{-1}|z\}-\mu\right]
\ee
where $\Theta=\int P(\mu,L)d\mu$ and $W\{\ldots|z\}$ is the length of the
primitive word on $G$ as a functional of the path on the complex plane
${\cal C}_1$. Conformal methods enable us to construct the topological
invariant $W$ for the group $G$.

To construct explicitly the topological invariants $W$ for general group
$G$ we go to the universal covering of double punctured plane ${\cal C}_1$.
This covering is isomorphic to the upper half--plane ${\cal H}=\{\zeta|\;
\zeta=\xi+i\lambda\; \lambda>0\}$.  The fundamental domain of the group
$G\{g_1,g_2\}$ has the form of the curvilinear triangle in ${\cal H}$. Each
fundamental domain represents the Riemann sheet corresponding to the fibre
bundle above ${\cal C}_1$. The universal covering space ${\cal Z}(\zeta)$
is the unit of all such Riemann sheets.

Coordinates of ends of the trajectory on ${\cal H}$ can be served as the
topological invariant for the path on double punctured plane ${\cal C}_1$ with
respect to the action of the group $G$. It
should be emphasized that this invariant is {\it complete} for our purposes.
In particular, coordinates of initial and final points of any
trajectory on ${\cal H}$ determine: (a) the coordinates of
corresponding points on ${\cal C}_1$; (b) the homotopy class of any path on
${\cal C}_1$: The paths on ${\cal H}$ are closed if
and only if $W\{g_1,g_2|z\}\equiv 1$, i.e. they belong to the trivial
homotopy class. After Fourier transform $P(\mu,L)= \frac{1}{2\pi}\int
\limits_{-\infty}^{\infty} e^{-ip\mu} P(q,L)dp$ the function $P(p,L)$
coincides with the Green function $P({\bf r}_0,{\bf r}={\bf r}_0,p,L)$ of
the non-stationary Schr\"odinger-like equation
\be \label{eq:10}
\frac{\partial}{\partial L}P({\bf r}_0,{\bf r},p,L)-\left(\frac{1}{2}{\bf
\nabla}-ip{\bf A}({\bf r})\right)^2 P({\bf r}_0,{\bf r},p,L)=
\delta(L)\delta({\bf r}-{\bf r}_0)
\ee
for the free particle motion in a "magnetic field" with the vector
potential ${\bf A}({\bf r})$ defined by the monodromy properties of the
group $G$; $p$ plays the role of a "charge" and the magnetic field is
transversal, i.e. $\mbox{rot}{\bf A}({\bf r})=0$. On the complex
plane ${\cal C}_1$ the field $A(z)$ is determined as $A(z)=d\zeta(z)/dz$.

Now let us construct the desired conformal mapping. The single-valued
function $z(\zeta)\equiv \zeta^{-1}(z)$ is defined in the fundamental
domain of the group $G$ on ${\cal H}$. The multivalued function
$\phi(\zeta)$ is determined as follows: (a) in the fundamental domain
$\phi(\zeta)=z(\zeta)$; (b) in all other domains on ${\cal H}$ the function $\phi(\zeta)$ is analytically continued through
the boundaries by means of the fractional transformations that are generated
by $G$.

Consider two basic contours $P_1$ and $P_2$ on ${\cal H}$ being the
conformal images of the contours $A_1(z)$ and $A_2(z)$ corresponding to the
generators $g_1$ and $g_2$ of $G$. The function $\phi(z)$ ($z\neq
\{z_1,z_2,\infty\}$) obeys the following transformations:
\be \label{eq:11}
\phi\left[z\stackrel{A_1}{\rightarrow} z\right]\rightarrow
\tilde{\phi}_1(z)= {{a_1\phi(z)+b_1}\over {c_1\phi(z)+d_1}}; \quad
\phi\left[z\stackrel{A_2}{\rightarrow} z\right]\rightarrow
\tilde{\phi}_2(z)= {{a_2\phi(z)+b_2}\over {c_2\phi(z)+d_2}}
\ee
where
\be \label{eq:12}
\left(\begin{array}{cc}a_1 & b_1 \\ c_1 & d_1 \end{array}\right) =
\hat{g}_1; \qquad \left(\begin{array}{cc}a_2 & b_2 \\ c_2 & d_2
\end{array}\right)=\hat{g}_2
\ee
are the matrices of basic substitutions of the group $G\{g_1,g_2\}$.

Assuming $\zeta(z)$ to be a ratio of two fundamental solutions, $u_1(z)$,
and, $u_2(z)$, of some second order differential equation with branching
points $\{z_1=0,\,z_2=1,\,z_3=\infty\}$, we conclude that the solutions
$u_1(z)$ and $u_2(z)$ undergo the linear transformations when the variable
$z$ moves along the contours $A_1(z)$ and $A_2(z)$:
\be \label{eq:12a}
A_1(z):\,\left(\begin{array}{c}\tilde{u}_1(z) \\ \tilde{u}_2(z)
\end{array}\right) = \hat{g}_1 \left(\begin{array}{c} u_1(z) \\ u_2(z)
\end{array}\right);\quad A_2(z):\,\left(\begin{array}{c}\tilde{u}_1(z) \\
\tilde{u}_2(z) \end{array}\right) = \hat{g}_2 \left(\begin{array}{c} u_1(z)
\\ u_2(z) \end{array}\right)
\ee

The problem of restoring the form of differential equation from the
monodromy matrices $g_1$ and $g_2$ of the group $G$ is known as
Riemann--Hilbert problem and has a long history \cite{venkov,bolib}.

Let us consider the special class of triangle groups---so-called Hecke
groups $H(h)$:
\be \label{eq:15}
T(h)=\left(\begin{array}{cc} 1 & h \\ 0 & 1 \end{array}\right);
\qquad S=\left(\begin{array}{cc} 0 & 1 \\ -1 & 0 \end{array}\right)
\ee
where $h=2\cos\frac{\pi}{q}$ and $q=3,4,5,6,\ldots$ (the case $h=1$
corresponds to $PSL(2,{\Z})$). The fundamental domain of the Hecke group
is the circular triangle $ABC$ with angles $\left\{0,\frac{\pi}{q},
\frac{\pi}{2} \right\}$ lying on ${\cal H}$.

The function $\disp\zeta(z,q)=u_1(z,q)/u_2(z,q)$ performs the
conformal mapping of Im$z>0$ with three punctures $(0,\, 1,\,
\infty)$ to the interior of the triangle $ABC$ on ${\cal H}$, where
$u_1(z,q)$ and $u_2(z,q)$ are the fundamental solutions of Picard-Fuchs
hypergeometric equation
\be \label{eq:16}
z(z-1) u''(z,q)+\left[\left(\frac{3}{2}-\frac{1}{q}\right)z-
\left(1-\frac{1}{q}\right)\right] u'(z,q)+\frac{1}{4}\left(
\frac{1}{2}-\frac{1}{q}\right)^2 u(z,q)=0
\ee

The former geometrical construction is evidently related to CFT.
Eq.(\ref{eq:16}) can be associated with the equation on the four-point
correlation function of some critical CFT. The question remains {\it
whether it is always possible to ajust the parameters of the corresponding
critical CFT to the coefficients of equations like} (\ref{eq:16}). In some
particular cases this question has positive answer and we describe briefly
the general construction.

Consider CFT characterized by the parameters $\left\{\Delta_1,\,
\Delta_2,\, \Delta_3=\Delta_1+\Delta_2-\Delta_{(12)},\, \delta\right\}$
which determine the conformal dimensions \cite{bpz}. Suppose that
$\varphi(z)$ is the primary field. The four--point correlation function
$\left<\varphi(z_1)\varphi(z_2) \varphi(z_3)\varphi(z) \right>$ satisfies
the partial differential equation, that follows from the conformal Ward
identity \cite{bpz,dots}. Introducing anharmonic ratio in the standard way,
we get
\be \label{eq:17}
\left[\frac{3}{2(2\delta+1)}\frac{d^2}{dz^2} + \left(\frac{1}{z} +
\frac{1}{z-1}\right)\frac{d}{dz} - \left(\frac{\Delta_1}{z^2} +
\frac{\Delta_2}{(z-1)^2} - \frac{\delta+\Delta_{(12)}}{z(z-1)} \right)
\right] \psi(z|0,1,\infty)=0
\ee
Recall, that $\delta\equiv\Delta_{(n,m)}=\frac{1}{24}(c-1)+ (\frac{1}{2}
\alpha_{+}n+\frac{1}{2}\alpha_{-}m)^2$ where $\alpha_{\pm}=
\frac{\sqrt{1-c}\pm\sqrt{25-c}}{\sqrt{24}}$; $(n,m)$ are positive integers
and $c$ is the central charge of corresponding Virasoro algebra.

The substitution $\disp\psi(z|0,1,\infty)=z^{-2b_{-}}(z-1)^{-2b_{+}}u(z)$
where $b_{\pm}=\frac{1}{6}(2\delta+1)-\frac{q\pm 1}{4q}$
allows us to rewrite (\ref{eq:17}) in the form of (\ref{eq:16}) by setting
\be \label{eq:18}
\left\{\begin{array}{l}
\disp \Delta_1=\frac{9-q^2+8\delta q^2-16\delta^2 q^2}{24(1+2\delta)q^2}
\medskip \\
\disp \Delta_2=\frac{9-18q+5q^2+32\delta q^2-64\delta^2 q^2}
{96(1+2\delta)q^2} \medskip \\
\disp \Delta_{(12)}=\frac{1}{2\delta+1}\left[\frac{9}{8q^2}-\frac{9}{8q}+
\frac{\delta}{2}\left(1-\frac{1}{q}\right)-\frac{10}{3}\delta^2+
\frac{1}{3}\right]
\end{array}
\right.
\ee
For $q=3$ we always have $\Delta_1=\Delta_2$ and one of possible solutions
(corresponding to $c=-2$) gives us $\disp \delta=-\frac{1}{8},\, \Delta_1=
\Delta_2=-\frac{5}{72},\, \Delta_{(12)}=-\frac{1}{72}$.

\section{Structure of Target Space of Braid Group $B_3$ and Random
Walk on $B_3$}

The braid group $B_n$ of $n$ strings has $n-1$ generators $\{\sigma_1,
\sigma_2,\ldots,\sigma_{n-1}\}$ with the following relations:
\be \label{eq:1}
\begin{array}{ll}
\sigma_i\sigma_{i+1}\sigma_i = \sigma_{i+1}\sigma_i\sigma_{i+1} & \qquad
(1\le i<n-1) \\ \sigma_i\sigma_j=\sigma_j\sigma_i & \qquad (|i-j|\ge 2) \\
\sigma_i\sigma_i^{-1}=\sigma_i^{-1}\sigma_i=e &
\end{array}
\ee
Let us mention that the {\it length} of the braid is the total number of
used letters, while the {\it minimal irreducible length} herafter referred
to as the "primitive word" is the shortest noncontractible length of a
particular braid which remains after applying all possible group relations
(\ref{eq:1}). Any braid corresponds to some knot or link. Hence, it is
feasible to use the braid group representation for
the construction of topological invariants of knots and links. However the
correspondence between braids and knots is not one--to--one and each knot
or link can be represented by infinite numbers of different braids.

The (right-hand) random walk (the random word) on the group $B_n$ with a
uniform distribution over generators $\{\sigma_1,\ldots,\sigma_{n-1},
\sigma_1^{-1},\ldots,\sigma_{n-1}^{-1}\}$ means that with the probability
$\frac{1}{2n-2}$ we add the element $\sigma_{\alpha_N}$ or
$\sigma_{\alpha_N}^{-1}$ to the given word of $N-1$ generators (letters)
from the right-hand side. The most attention should be paid to the
question: what is the probability $P(\mu,N)$ to find the $\mu$--letter
primitive word in the $N$--letter random word in the group $B_n$. In
\cite{neve} it is conjectured that $P(\mu,N)$ obeys the following
asymptotics \be \label{eq:distrib} P(\mu,N)\propto
\frac{\mu}{N^{3/2}}\exp\left(-a(n)N+b(n)\mu-\frac{\mu^2}{c(n)N}\right) \ee
where $a(n),\;b(n),\;c(n)$ are numerical constants depending on $n$ only.

The group $B_3$ can be represented by $PGL(2,{\R})$--matrices. To be more
specific, the braid generators $\sigma_1$ and $\sigma_2$ in the Burau
representation \cite{birman} read
\be \label{eq:2}
\sigma_1=\left(\begin{array}{cc} -t & 1 \\ 0 & 1 \end{array} \right);
\qquad \sigma_2=\left(\begin{array}{cc} 1 & 0 \\ t & -t
\end{array}\right)
\ee
where $t$ is the "spectral parameter". For $t=-1$ the matrices $\sigma_1$
and $\sigma_2$ generate $PSL(2,{\Z})$ such that $B_3$ is its central extention.

We define $\tilde{\sigma}_1=\sigma_1,\; \tilde{\sigma}_2= \sigma_2$ (at
$t=-1$) and rewrite the generators of the modular group $PSL(2,{\Z})$ in
the canonical form using the matrices $S,T$ (compare to (\ref{eq:15})). The
braiding relation $\tilde{\sigma}_1\tilde{\sigma}_2\tilde{\sigma}_1=
\tilde{\sigma}_2\tilde{\sigma}_1\tilde{\sigma}_2$ in the $\{S,
T\}$--representation takes the form $S^2TS^{-2}T^{-1}=1$ (we also have $S^4
=(ST)^3=1$).

The group $PSL(2,{\Z})$ is a discrete subgroup of $PSL(2,\R)$. The
fundamental domain of $PSL(2,\Z)$ is a circular triangle with angles
$\left\{0,\frac{\pi}{3}, \frac{\pi}{2}\right\}$ located in ${\cal H}$. The
group $PSL(2,\Z)$ is completely defined by its
basic substitutions: (i) $S:\,\zeta\to -1/\zeta$; (ii) $T:\,\zeta \to
\zeta+1$. Let us choose an arbitrary element $\zeta_0$ from the fundamental
domain and construct a graph, $C(\Gamma)$ connecting the neighboring images
of the initial element $\zeta_0$ obtained under successive action of the
generators from the set $\{S,T,S^{-1},T^{-1}\}$. The graph $C(\Gamma)$ is
shown in the Fig.\ref{fig:1}a. It will be seen that although this graph has
the local cycles, its "backbone" has a Cayley tree ("free group") structure.
Thus, the problem of limit distribution of a random walks on the modular
group can be related to the problem of random walks on the
graph $C(\Gamma)$. Hence, we immediately can compute the probability
$P(\mu,N)$ that the $N$-step random walk along the graph $C(\Gamma)$ which
starts from the origin ends in some vertex of the cell at the
distance of $\mu$ steps along the backbone. This distribution has the form
of Eq.(\ref{eq:distrib}).

\begin{figure}[t]
\vspace*{0.5cm}
\hspace{1.7cm}\special{em:graph mon1_fig.pcx}
\vspace*{7cm}
\caption{\small The graph corresponding to the structure of:
(a) the modular group $PSL(2,{\Z})$ (i.e. the Hecke group with $q=3$) and
(b) the Hecke group with $q=4$.}
\label{fig:1}
\end{figure}

Now we can construct the "target space" of the braid group $B_3$. Introduce
the normalized generators of the group $B_3$: $||\sigma_1||=\tau(t)$;
$||\sigma_2||=\tau^{-1}(t)s(t)\tau^{-1}(t)$, where $||\sigma_1(t)||= (\det
\sigma_1)^{-1}\sigma_1$ and $||\sigma_2(t)||= (\det \sigma_2)^{-1}
\sigma_2$. By $\tau(t)$ and $s(t)$ we denote the generators of the
"$t$-deformed" group $PSL_t(2,\Z)$. The group $PSL_t(2,{\Z})$ keeps
invariable the relations of $PSL(2,{\Z})$, namely, we have
$(\tau(t)s(t))^3= s^4(t)= \tau(t)s^2(t)\tau^{-1}(t) s^{-2}(t)=1$. Hence, we
ultimately come to the conclusion that the graphs $C(\Gamma_t)$ (for
$PSL_t(2,{\Z})$) and $C(\Gamma)$ are topologically equivalent. However the
metric properties of $C(\Gamma_t)$ and $C(\Gamma)$ are different because of
different embeddings of $PSL_t(2,\Z)$ and $PSL(2,\Z)$ into the upper
half--plane. The additional benefit of this construction is that we can
represent the graph of $B_3$ on the graph $C(\Gamma)$ if we supply in each
center of the cell the "magnetic pole" with the quantity of the flux equal
to $t^6$. Fig.\ref{fig:1}b represents the graph corresponding to Hecke
group with $q=4$.

To combine the monodromy representation of the Hecke group $H(h)$ with
the topological properties of knots on the basis of braid group
representation, consider the space of functions $V=\left\{f_{ij}\right\}$
where
\be \label{eq:19}
f_{ij}\left(x_1,x_2,\ldots,x_n\right)=\int_{x_i}^{x_j}
(z-x_1)^{-\mu_1}\ldots (z-x_n)^{-\mu_n}dz
\ee
$\mu_1,\,\ldots,\,\mu_n$ are the complex numbers and path integration is
selected in the space $CP^1\backslash \left\{x_1,\ldots,\, x_n \right\}$.
The group $B_n$ acts on $f_{ij}$ if $\mu_i=\mu_j$ but the group of colored
braids $C_n$ acts on the general linear span of the multivalued functions
$\left\{f_{ij}:\; 0\le i<j\le n-1 \right\}$.

In particular case of $n=3$ we study the integral
$$
\int \omega=\int z^{-\mu_1}(z-1)^{-\mu_2}(z-x)^{-\mu_3}dz
$$
Due to the classical result of H.A. Schwarz \cite{most} we deduce that for
any set of points $x_1,\,x_2 \notin \{0,\,1,\,\infty,\,x\}$ the integral
$\disp\int_{x_1}^{x_2}\omega$ is the multivalued function of $x$ satysfying
the general hypergeometric equation
$$
z(z-1)u''(z)+\{(\alpha+\beta+1)z-\gamma\}u'(z)+\alpha\beta u(z)=0
$$
where $\alpha,\,\beta,\,\gamma$ are arbitrary real constants related to
$\mu_i$. The analysis
of this equation shows that the set of all monodromies arising by moving
one puncture ($x$, for instance) around other ones $(0,\,1,\,\infty)$
generates the monodromy group $C_4(CP^1)$ on $V$. This group is discrete
for some special constraint on exponents $\{\mu_i\}$ only (see
\cite{most}). Denote these monodromies by $A(x|0),\,A(x|1),\,A(x|\infty)=
[A(x|0)A(x|1)]^{-1}$ (compare to (\ref{eq:12a})).

Let us briefly discuss the procedure of constructing the invariants of knots
generated by the representations of $C_4$. In the Jones' theory of polynomial
invariants it has been pointed out the remarkable connection between
representations of II$_1$--factors and knot invariants \cite{jones}. In
particular it was proved that if the index value of factor II$_1$ is less
than 4, there exists an integer $q\ge 3$ with index value $r=4\cos^2
\frac{\pi}{q}$. The II$_1$--factor is hyperfinite.

To each element of the group $C_4(CP^1)$ we associate the word $W$ represented
by the monodromy matrices $\hat{g}_1$ and $\hat{g}_2$. Any braid (a)
generates some knot and (b) completely determines the monodromy matrix
$M^{(N)}=\prod_{i=1}^N g_{\alpha_i}$, where $g_{\alpha_i}= \{g_1,\,g_2,\,
g_1^{-1},\, g_2^{-1}\}$ ($i=1,2,\ldots, N$) and $N$ is the total number
generators used in the word $W$. Thus, tracing of $M^{(N)}$ gives the
invariant of a knot. Using the classification of discrete subgroups of the
$PSL(2,{\R})$ with two generators \cite{gil} (which includes also the
Hecke and other triangle groups) we can expand our construction to the
general case.

\section*{Concluding Remarks}

1. Comparing the target spaces of the modular ($PSL(2,{\Z})$) and free
($\Gamma_2$) groups we see that the branching $f_{\rm eff}$ of the graph
(i.e. the effective "curvature" of the corresponding hyperbolic space, in which
the graph could be embedded) varies for different $q$. In particular,
$f_{\rm eff}$ is changing from $f_{\rm eff}=3$ for $PSL(2,{\Z})$ to $f_{\rm
eff}=4$ for $\Gamma_2$.

2. Let us stress the that the index value $r$ of II$_1$--type factors,
is equal to $h^2$ where $h$ is the Hecke group parameter. Thus, the braids
connected with the representations of II$_1$--type factors can be realized
as monodromies of Hecke groups and corresponding knot invariants could be
produced by tracing of monodromy matrices.

3. Topological properties of knots generated by $C_4$--group and its
monodromy representations are related to each other by the following
construction. Take the manifold $H/\Gamma$ and consider its
compactification $\overline {H/\Gamma}$. The genus of $\overline{H/\Gamma}$
can be computed using the Hurwitz's formula. Each closed path $\gamma$ on
$\overline{H/\Gamma}$ coresponds to some closed $B_3$-- or $C_4$--braid.
The path can be expanded through the (non--trivial) conjugacy classes of
$\Gamma$ which in turn determine the group $C_4(CP^1)$ due to the
uniqueness of the monodromy representation.

In each conjugacy class of the group $\Gamma$ lies the shortest closed
path (geodesics). Since the length of geodesics are completely determined
by the spectrum of the Laplacian $\Delta$ (by the "partition function") on
the corresponding Riemann surface, the connection between the knot
invariants and the partition function of $\Delta$ should exist. The natural
formula, linking the spectral properties of the Laplacian on $\overline
{H/\Gamma}$ with the conjugacy classes (of hyperbolic, elliptic and
parabolic types) of $\Gamma$ is the Selberg trace formula. Hence, we claim
it is possible to express the invariants of knots represented by $B_3$--
$C_4$--braids via Selberg trace formula. More details will be discussed in
the forthcomming publication.
\medskip

\noindent{\bf Acknowledgements}

We thank L. Chayes and S. Hirson for the help in final preparation
of the paper. This research is partially supported by  the RFBR grant
96--01--00744.


\newpage

\begin{thebibliography}{99}
\bibitem{wil} F. Wilczeck, {\it Fractional Statistics and Anyon
Superconductivity} (WSPC: Singapore, 1990)
\bibitem{mon} M. Monastyrsky, {\it Topology of Gauge Fields and Condensed
Matter} (Plenum: NY, 1993)
\bibitem{nech} S. Nechaev, {\it Statistics of Knots and Entangled Random
Walks} (WSPC: Singapore, 1996)
\bibitem{grosb} A.Yu. Grosberg, A.R. Khokhlov, {\it Statistical Physics of
Macromolecules} (AIP Press: NY, 1994)
\bibitem{int} {\it Integrable Models and Strings}, Lect. Notes Phys., 436,
(Springer: Heidelberg, 1994)
\bibitem{witten} E. Witten, Comm. Math. Phys., 121 (1989), 351
\bibitem{birman} J. Birman, {\it Knots, Links and Mapping Class Groups},
Ann. Math. Stud., 82, (Princeton Univ. Press: Princeton, NJ, 1976)
\bibitem{venkov} A.B. Venkov, Preprint LOMI P/2/86
\bibitem{bolib} D. V. Anosov, A.A. Bolibruch, {\it Riemann--Hilbert
problem}, (Vieweg: Braunschweig, 1994)
\bibitem{neve} S. Nechaev, A. Vershik, J. Phys. (A): Math. \& Gen., 27
(1994), 2289
\bibitem{bpz} A.A. Belavin, A.M. Polyakov, A.B. Zamolodchikov, Nucl.
Phys.(B), 241 (1984), 333
\bibitem{dots} Vl.S.Dotsenko, Nucl.Phys. B-235 [FS11] (1984), 54
\bibitem{most} G.D. Mostov, Bull. AMS, 16 (1987), 225
\bibitem{gil} J. Gilman, Memoirs AMS, 1996 (in press)
\bibitem{jones} V.F.R. Jones, {\it Subfactors and Knots}, CBMS No. 80,
(AMS: Providence, 1991)
\end{thebibliography}
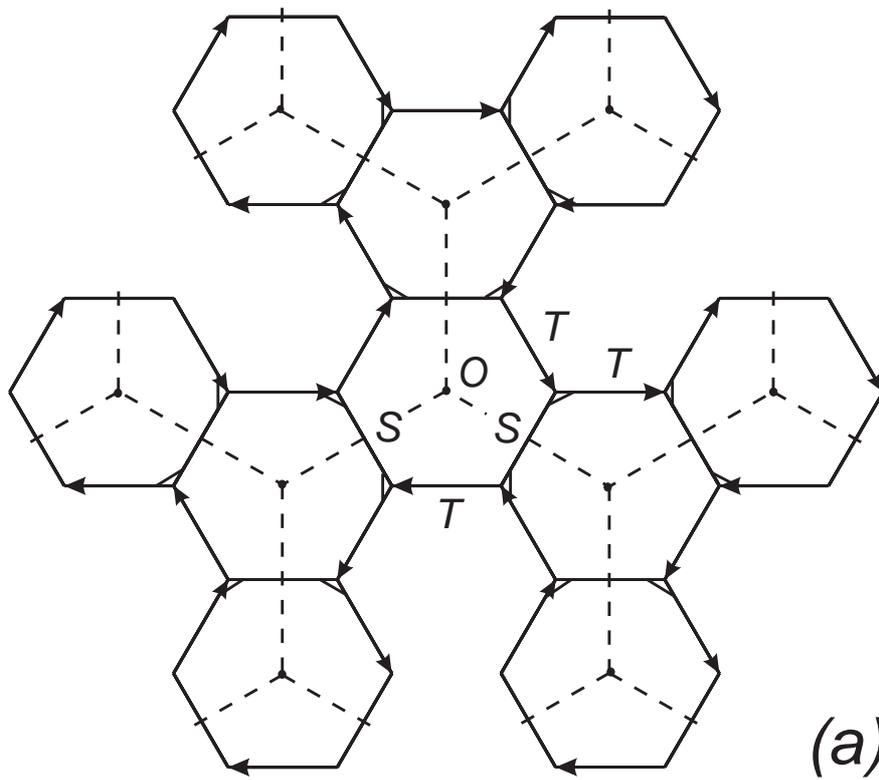
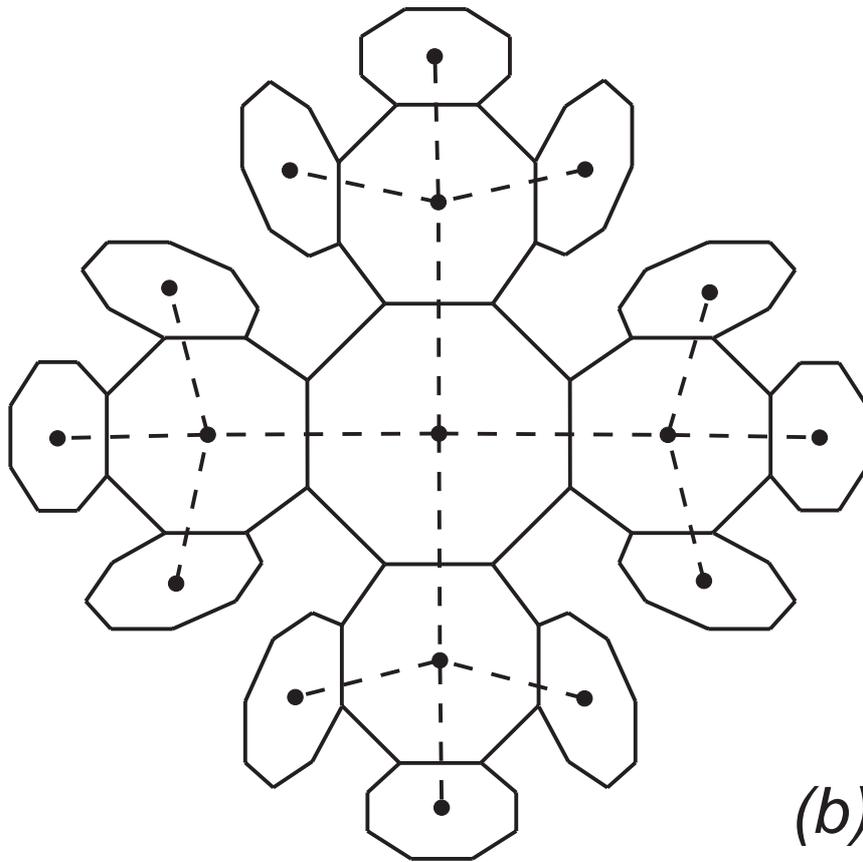
\end{document}